\documentclass[aps,prl,preprint,groupedaddress]{revtex4-1}
\usepackage{graphicx}
\graphicspath{ {./Figures/} }
\usepackage{epstopdf}
\usepackage{float}

\begin{document}

%\preprint{}

\title{Gate-tunable band structure of the LaAlO$_{3}$-SrTiO$_{3}$ interface}


\author{A.E.M. Smink, J.C. de Boer, M.P. Stehno, A. Brinkman, W.G. van der Wiel, and H. Hilgenkamp}
%\thanks{}
\affiliation{MESA+ Institute for Nanotechnology, University of Twente, P.O. Box 217, 7500 AE Enschede, The Netherlands}


\date{\today}

\begin{abstract}
The 2-dimensional electron system at the interface between LaAlO$_{3}$ and SrTiO$_{3}$ has several unique properties that can be tuned by an externally applied gate voltage. In this work, we show that this gate-tunability extends to the effective band structure of the system. We combine a magnetotransport study on top-gated Hall bars with self-consistent Schr\"odinger-Poisson calculations and observe a Lifshitz transition at a density of $2.9\times10^{13}$ cm$^{-2}$. Above the transition, the carrier density of one of the conducting bands decreases with increasing gate voltage. This surprising decrease is accurately reproduced in the calculations if electronic correlations are included. These results provide a clear, intuitive picture of the physics governing the electronic structure at complex oxide interfaces.
\end{abstract}

%\keywords{}

\maketitle


%\section{Introduction}
The two-dimensional electron system (2DES) at the interface between the band insulators LaAlO$_{3}$ (LAO) and SrTiO$_{3}$ (STO) displays many intriguing phenomena, which may be harnessed for novel electronic devices \cite{ohtomo_high-mobility_2004, mannhart_oxide_2010, hwang_emergent_2012, xie_tuning_2013, hilgenkamp_novel_2013}. The discovery of superconductivity \cite{reyren_superconducting_2007}, magnetic signatures \cite{brinkman_magnetic_2007, ariando_electronic_2011, lee_titanium_2013, banerjee_ferromagnetic_2013}, and their apparent coexistence \cite{li_coexistence_2011} sparked growing interest in this material system. These properties can be tuned by varying parameters during growth \cite{brinkman_magnetic_2007, xie_tuning_2013}, as well as by an externally applied electric field after growth \cite{thiel_tunable_2006}. Using this field-effect, control of superconductivity \cite{caviglia_electric_2008, richter_interface_2013, eerkes_modulation_2013, ben_shalom_tuning_2010}, of spin-orbit coupling \cite{caviglia_tunable_2010, ben_shalom_tuning_2010, stornaiuolo_weak_2014, hurand_field-effect_2015}, and of carrier mobility \cite{bell_dominant_2009, hosoda_transistor_2013} have been reported. Recent progress on local control of superconductivity \cite{eerkes_modulation_2013} opened a route towards electrically controlled oxide Josephson junctions \cite{goswami_nanoscale_2015, goswami_quantum_2016}, opening new opportunities for superconducting electronic devices. Because these phenomena are related to the interfacial band structure, a fundamental understanding of the band structure is vital for the understanding of these phenomena and their exploitation in electronic devices.

At the interface, the conduction band of STO is bent down and crosses the Fermi level \cite{popovic_origin_2008}. The origin of this band banding is still an open question \cite{nakagawa_why_2006, siemons_origin_2007, yu_polarity-induced_2014}. It creates a potential well, confining the carriers to a few nm in the out-of-plane direction \cite{basletic_mapping_2008, sing_profiling_2009, reyren_anisotropy_2009, copie_towards_2009}. In the well, the effective band structure is formed by the Ti $t_{2g}$ orbitals. For interfaces grown along the [001] direction, the $d_{xy}$ bands lie below the $d_{yz,xz}$ bands in energy \cite{salluzzo_orbital_2009}. To properly describe the band structure based on these observations, the splitting in energy between these bands $\Delta E$ and their effective masses $m^*$ are key parameters. Especially for $\Delta E$, a large variation in values has been reported for both LAO-STO interfaces \cite{salluzzo_orbital_2009, mccollam_quantum_2014} and oxygen-deficient SrTiO$_{3-\delta}$ \cite{meevasana_creation_2011, santander-syro_two-dimensional_2011, plumb_mixed_2014}.

In transport experiments, one can move the position of the Fermi level in a band structure by electrostatic gating and extract information about the physics governing conduction in the material. By back-gating the interface through the STO substrate, an additional conduction channel was observed to emerge above a carrier density of $(1.7\pm0.1)\times10^{13}$ cm$^{-2}$ \cite{joshua_universal_2012}. This observation was linked to tuning the Fermi level across the bottom of the $d_{yz,xz}$ bands, where additional electron pockets become available for conduction. This changes the topology of the Fermi surface, which is the characteristic feature of a Lifshitz transition \cite{yamaji_quantum_2006}. The extracted value for $\Delta E$ in a  single-subband approximation, $\sim 58$ meV, is close to the value obtained from X-ray absorption measurements \cite{salluzzo_orbital_2009} and a universal value for the carrier density at the Lifshitz point was reported in Ref.\ \cite{joshua_universal_2012}.

The underlying model implied a set of band structure parameters that may not apply in general. Subsequent experiments using different gating setups were not able to reproduce the Lifshitz transition scenario with the universal value for the carrier density \cite{hosoda_transistor_2013, eerkes_modulation_2013, hurand_field-effect_2015, liu_magneto-transport_2015}. In these studies, a carrier density up to $2.5\times10^{13}$ cm$^{-2}$ was reported, i.e. well above the critical density mentioned above. The absence of the Lifshitz transition in these studies suggests that the Lifshitz density is not an intrinsic property of the interface, but depends crucially on the electrostatic boundary conditions among other parameters.

In this Letter, we report top-gating experiments of LAO-STO interface devices over a wide gate voltage range with record low gate-leakage currents. From two-band-fits on the magnetotransport data, we extract the evolution of the $d_{xy}$ and $d_{yz,xz}$ carrier densities as function of top-gate voltage. By self-consistent Schr\"odinger-Poisson calculations, we provide a simple and intuitive picture of the band structure of the 2DES, which crucially depends on out-of-plane electrostatics. This approach naturally explains our observation of a Lifshitz transition at a carrier density of $(2.9\pm0.1)\times10^{13}$ cm$^{-2}$ \--- almost twice the value reported for back-gating \--- and the large variation of values reported in literature. Above the transition, we observe a reduction of the $d_{xy}$ carrier density, which can be attributed to electron-electron interactions.

\begin{figure}
\includegraphics{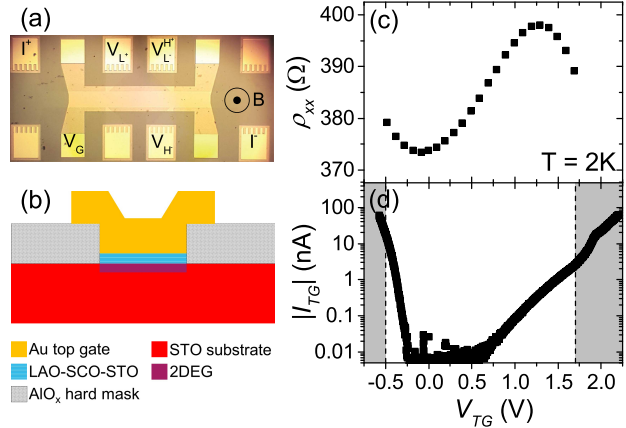}
\caption{Device layout and basic electrical properties. (a) Optical micrograph of a 75 $\mu$m wide Hall bar with electrical connections and magnetic field direction indicated. (b) Schematic cross-section of device. (c) Sheet resistance and (d) Absolute gate leakage current as function of top-gate voltage at $T$ = 2 K and $B$ = 0 T. The shaded area is excluded for further analysis due to conditions discussed in the main text.}
\label{device}
\end{figure}

%\section{Methods}
To study the top-gate dependence of the transport parameters, several Hall bars were fabricated on SrTiO$_{3}$ substrates, which were terminated on the TiO$_{2}$-planes by etching with buffered hydrofluoric acid (BHF) \cite{koster_quasi-ideal_1998}. Subsequently, a $\sim$20 nm thick amorphous AlO$_{x}$ hard mask was deposited and Hall bar structures were defined through photolithography and etching in a base \cite{banerjee_direct_2012}. These Hall bars are 30 - 100 $\mu$m wide and 400 $\mu$m long. After structuring, 10 unit cells (u.c.) of LaAlO$_{3}$ were deposited using pulsed laser deposition at 850$^{\circ}$C and an oxygen pressure of 5$\times$10$^{-5}$ mbar, with a laser fluence of 1.3 J cm$^{-2}$ and a laser spot size of 1.76 mm$^{2}$. The target-substrate distance was 45 mm. To suppress formation of oxygen vacancies, the film was capped \textit{in situ} with 1 u.c. of SrCuO$_{2}$ and 2 u.c. of SrTiO$_{3}$, at 600$^{\circ}$C and in 6$\times$10$^{-2}$ mbar of oxygen \cite{huijben_defect_2013}. The sample was then cooled down to room temperature at 10$^{\circ}$C/min in deposition pressure and transferred \textit{ex situ} to a sputtering chamber. There, a 30 nm thick Au layer was deposited at a low rate of $\sim$1 nm/min in 8$\times$10$^{-2}$ mbar of argon gas. Au/Ti contact pads were defined by another sputtering step, after which the gate electrodes were defined by etching the excess Au in a buffered KI solution. A photograph and a schematic of the device are shown in Figs.\ \ref{device}(a)-(b). The top-gate voltage $V_{TG}$ was applied between the gate electrode, $V_G$, and the source, $I^-$. The transport measurements were performed in a Physical Property Measurement System, and the gate leakage current characteristics of Fig.\ \ref{device}(d) were measured with a Source Measure Unit. We measured several devices on two different samples showing similar behavior. In the following, we discuss results for one of these devices.

%\section{Results}
The gate leakage current, $I_{TG}(V_{TG})$, depicted in Fig.\ \ref{device}(d) is very small compared to values reported in literature \cite{hosoda_transistor_2013,eerkes_modulation_2013, woltmann_field-effect_2015}, given the size of the gated area: $\sim$$4\times10^4$ $\mu$m$^{2}$. This indicates that the dielectric of our samples is an excellent insulator, which can be ascribed to the gentle metal deposition and/or to the SrCuO$_2$ capping. The effect of the latter would be enhanced oxygen uptake \cite{huijben_defect_2013} and a small defect density in the overlayer. For the transport measurements, we chose the gate voltage range between -0.5 V and +1.7 V. In this range, two requirements are met: the gate leakage current is maximally 1\% of the excitation current, and no dielectric breakdown is observed. We interpret the upturn of $I_{TG}$ above $V_{TG} = +1.7$ V as an onset of breakdown, as the corresponding electric field is $\gtrsim 3.5$ MV cm$^{-1}$.

\begin{figure*}
\includegraphics{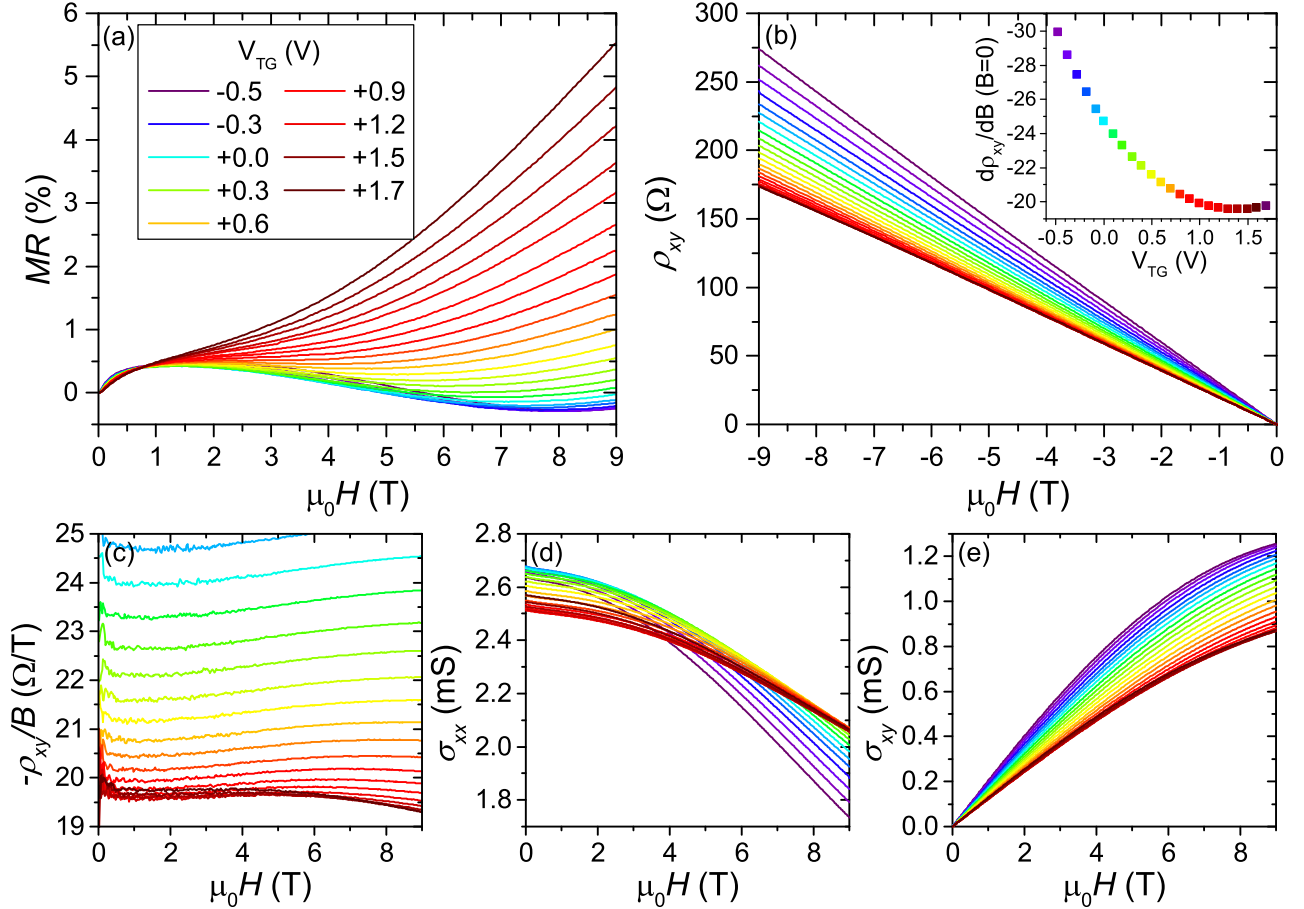}
\caption{Magnetotransport curves after symmetrization at $T$ = 2 K. (a) Magnetoresistance, normalized to zero-field value, showing positive magnetoresistance for $V_{TG}\gtrsim$ +0.3 V. The legend applies to all graphs. (b) Hall resistance showing an emerging slight nonlinearity around $V_{TG}\sim$ +1.2 V. The inset shows the zero-field slope. (c) Hall coefficient, where the nonlinearity in the Hall resistance is more visible as the signal starts turning downwards at higher field with increasing gate voltage. (d) Longitudinal conductivity showing a change in behavior around $V_{TG}$ = +0.5 V. (e) Hall conductivity displaying an increasing low-field slope for $V_{TG}\gtrsim$ +1.0 V.}
\label{mr}
\end{figure*}

Within this range, the sheet resistance depends on the top-gate voltage in a non-trivial way, as depicted in Fig.\ \ref{device}(c). The behavior on the left hand side ($V_{TG} \lesssim 0.5$ V) can be explained by opposite trends in the density and mobility of the charge carriers with gate voltage \cite{hosoda_transistor_2013}, whereas the right-hand side requires additional explanation. To shed light on the origin of this unusual gate-dependence of the resistivity, magnetoresistance measurements were performed in 100 mV gate voltage steps. The results are depicted in Figs.\ \ref{mr}(a)-(c). Here, the magnetoresistance is defined relative to the zero-field value, \(MR=[\rho_{xx}(B)/\rho_{xx}(0)-1]\times100\%\). Besides low-field signatures of weak antilocalization \cite{caviglia_tunable_2010}, we observe the characteristic features attributed to a Lifshitz transition in the band structure of the LAO-STO interface \cite{joshua_universal_2012}: the emerging positive magnetoresistance, nonlinear Hall signal and an upturn of the low-field Hall slope. However, the characteristic changes all occur at different gate voltages.

\begin{figure}
\includegraphics{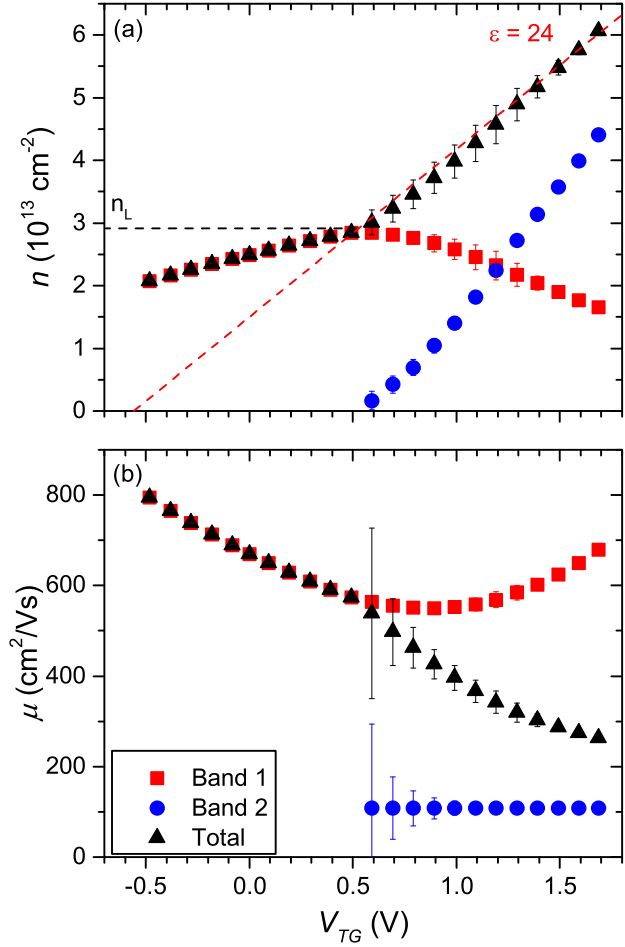}
\caption{Transport parameters, as extracted from fitting by the two-band model, as function of top-gate voltage, at $T$ = 2 K (a) Carrier density per carrier type. The black dashed line indicates the Lifshitz carrier density. The red line is the slope of the carrier density calculated from a parallel-plate capacitor model with $\epsilon=24$ and $d = 5$ nm. (b) Carrier mobility, where $\mu_{tot}$ is calculated as the weighted average of the two bands.}
\label{nandmu}
\end{figure}

To extract the carrier density as function of gate voltage, we inverted the 2D resistivity matrix to obtain the conductivity in longitudinal ($\sigma_{xx}$) and transverse ($\sigma_{xy}$) direction as a function of magnetic field, plotted in Figs.\ \ref{mr}(d)-(e). For every top-gate voltage, these curves were made to fit simultaneously to a two-band conduction model: 
\begin{equation}
\sigma_{xx}=e\displaystyle\sum_{i=1,2}\frac{n_{i}\mu_{i}}{1+(\mu_{i}B)^2}; \hspace{15pt}\sigma_{xy}=eB\displaystyle\sum_{i=1,2}\frac{n_{i}\mu^2_{i}}{1+(\mu_{i}B)^2},
\end{equation}
where $n_i$ is the carrier density and $\mu_i$ the mobility of the $i$-th band, and $B$ is the magnetic field. For fitting through the Lifshitz transition, we assume continuity of $n_{1}$ and $n_{2}$. This corresponds to requiring the Fermi surface area to be continuous as function of chemical potential. As shown in Fig.\ \ref{nandmu}(a), this condition can be met by assuming a lower limit for the mobility of the second band, $\mu_2$. This avoids $n_2$ to diverge just above the transition. The error bars on $\mu_{2}$ represent standard errors from the least-squares fit and correlate to the ratio of conductivity of the separate bands, $n_{1}\mu_{1}/n_{2}\mu_{2}$. The error bars on $n_{1}$ and $\mu_{1}$ are calculated by both the fitting error and the spread of $n_{1}$ values for fits with and without the lower limit on $\mu_2$.

%\section{Discussion}
The most notable observation in Fig.\ \ref{nandmu} is the emergence of a second mobile carrier type around a carrier density of $(2.9\pm0.1)\times10^{13}$ cm$^{-2}$, which we interpret as the Lifshitz density $n_L$ of this sample. Since this Lifshitz density is almost twice the value of Ref.\ \cite{joshua_universal_2012}, $\Delta E$ is evidently larger in this case. This means that the effective band structure in the two cases must differ. Below, we discuss this result qualitatively using a Schr\"odinger-Poisson (SP) approach \cite{stern_self-consistent_1972, gariglio_electron_2015}.

In SP calculations, the wave function and confining potential are calculated self-consistently by solving the coupled Schr\"odinger and Poisson equations. To model the LAO-STO interface, we imply an infinite energy barrier on the LAO side of the interface, and fix the electric field strengths at the interface ($z=0^+$) and deep in the STO ($z=\infty$). Without applied gate voltage, the total charge at the interface determines the interfacial electric field strength, and the field strength deep inside the STO is zero. An applied gate voltage has the primary effect to dope more carriers through capacitive coupling. This results in a steeper potential well, or a larger electric field strength. A steeper gradient in potential increases the energy of a state in the well. This depends inversely on the state's effective mass in the confinement direction, which is large for $d_{xy}$ states and small for $d_{yz,xz}$ states. The result is that $d_{xy}$ states lie below the $d_{yz,xz}$ states in energy, in accordance with experimental results \cite{salluzzo_orbital_2009}. This also entails that the splitting in energy $\Delta E$ depends on the gradient in potential. In the following, as only $d_{xy}$ states are available below $n_L$, the carriers labeled $n_1$ are of $d_{xy}$ character and $n_2$ represents the $d_{yz,xz}$ states. 

By simple electrostatic reasoning, a positive top-gate voltage increases the electric field strength at $z=0^+$, and a positive back-gate voltage decreases the electric field strength at $z=\infty$ \cite{scopigno_phase_2016}. Therefore, a positive top-gate (back-gate) voltage enhances (weakens) the effect of electrostatic doping on the confining potential gradient. 
As $\Delta E$ depends on the gradient of the confining potential, a positive top-gate voltage increases $\Delta E$, and a positive back-gate voltage decreases it. This is in line with the observation of the higher Lifshitz density in our experiment.

Above the Lifshitz transition, we observe the $d_{xy}$ carrier density to decrease with increasing gate voltage. Such a decrease is incompatible with a model requiring a fixed electronic band structure, as raising the Fermi energy should always increase the number of available conduction states up to the point where the band is full. This is not the case here. Instead, it appears that the carriers redistribute to the $d_{yz,xz}$ bands. Below, we show that this behavior follows naturally if electron-electron interactions are taken into consideration, as first introduced by Maniv \textit{et al.}\ \cite{maniv_strong_2015}.

For the numerical calculations, we follow the method of Ref.\ \cite{gariglio_electron_2015} and assume that the potential well is formed by the mobile carriers $n_m$ and a bound background charge $n_{b}$ spread homogeneously across a thickness of 100 nm. An important feature of STO is the electric-field-dependence of its dielectric constant, which thus varies over the interface region. By using the dependence provided in Ref.\ \cite{gariglio_electron_2015}, it is evaluated for each iteration step of the self-consistent calculation. In the model, we vary $n_m$ and $n_b$ to study the occupancy of the $d_{xy}$ and $d_{yz,xz}$ states. Importantly, a change in $n_{b}$ influences the Lifshitz density as this alters the shape of the potential well. By choosing $n_{b} = 5.75\times10^{18}$ cm$^{-3}$, we reproduce the experimentally found Lifshitz density. The effective masses are taken as $m_l = 0.7$ $m_e$ and $m_h=14$ $m_e$ \cite{santander-syro_two-dimensional_2011}. We note that a large range of effective masses has been reported in literature, e.g.\ \cite{zhong_theory_2013, mccollam_quantum_2014}, but that we can reproduce the top-gate Lifshitz density for this range of effective masses by choosing a different background charge density.

\begin{figure}
\includegraphics{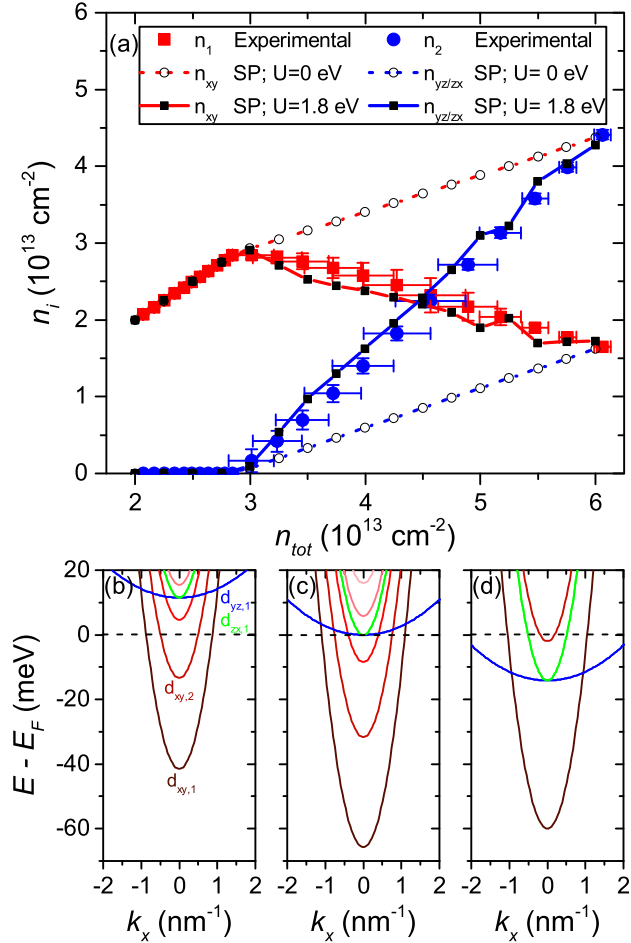}
\caption{Results of the self-consistent Schrodinger-Poisson calculations. (a) Calculated and measured carrier density per band versus the total carrier density. The symbols with error bars represent the experimental values; the symbols with connecting lines (as guide to the eye) correspond to calculated values by the Schr\"odinger-Poisson model as described in the text. Dashed lines (open symbols) represent calculations without electron-electron interactions, solid lines are calculated with $U=1.8$ eV.  (b)-(d) Calculated band structures including electron-electron interactions for (b) $n_{tot}\ll n_L$, (c) $n_{tot} = n_L$, and (d) $n_{tot}\gg n_L$. The horizontal dashed line represents the Fermi energy and the color legend in (b) applies to all panels.}
\label{SPcalc}
\end{figure}

The results of the SP calculation are given by the open symbols and dashed lines in Fig.\ \ref{SPcalc}(a), which clearly does not reproduce the decrease of $n_1$. The redistribution of carriers to the $d_{yz,xz}$ bands can thus not be explained by interaction of the potential well and the band structure alone, but has to be mediated by an effect not yet included in the calculations. As recently reported by Maniv \textit{et al.}\ \cite{maniv_strong_2015}, electronic correlations can mediate such a redistribution of carriers. 

We model these correlations as Hubbard-like electron-electron interactions, with for each band $i$ an interaction term \(E_{int}^i=\sum\nolimits_{j=1}^nU\left(1-\delta_{ij}/2\right)N_{j}\). Here, $N_{j}$ is the 2D electron density per unit cell of band $j$, $\delta_{ij}$ is the Kronecker-delta, and $U$ is the phenomenological screened Coulomb interaction strength between bands, which we take equal for all bands for simplicity. The Kronecker delta term is included to avoid unphysical self-interactions. We include this interaction term to the self-consistent band occupancy calculation in the SP model. 

Using $U = 1.8$ eV, the evolution of $n_1$ and $n_2$ is calculated as the solid lines in Fig.\ \ref{SPcalc}(b), closely resembling the experimental data. The resulting band structures for $n\ll n_L$, $n=n_L$, and $n\gg n_L$ are given in Figs.\ \ref{SPcalc}(b)-(d), respectively. From these calculations, it becomes clear that the net effect of this electron-electron interaction is to shift bands with low density of states (DOS) up in energy with respect to the bands with high DOS, once the latter are occupied.

%\section{Conclusion}
In summary, we have observed a Lifshitz transition in the band structure of the LaAlO$_{3}$-SrTiO$_{3}$ interface at a carrier density of $(2.9\pm0.1)\times10^{13}$ cm$^{-2}$. Above the transition, the density of $d_{xy}$-type charge decreases with increasing gate voltage. To investigate this behavior, we performed self-consistent band structure calculations by solving the coupled Schr\"odinger and Poisson equations. The obtained band occupations are in excellent agreement with the experimental data when an electron-electron interaction term is included in the model. This term mediates a shift in $d_{xy}$ subbands when $d_{yz,xz}$ subbands become occupied. 

Based on these observations, we conclude that electrostatics and electronic correlations are equally important factors governing the band structure of SrTiO$_3$-based 2-dimensional electron systems. The growth conditions of the polar oxide, dopants and oxygen deficiency in the SrTiO$_3$ substrate, and the presence of a metallic top-gate electrode determine the shape of the potential well at low carrier concentrations. Once the heavier $d_{yz,xz}$ bands are populated, electron-electron interactions wield influence on the relative band occupations. Therefore, the relative band offsets of the 3d-bands and the carrier density of the Lifshitz transition associated with the occupation of the heavier bands of out-of-plane orbitals should not be considered fixed, but they evolve with the shape of the potential well.   

An improved understanding of the band structure at complex oxide interfaces is an important step in the direction of a universal framework that explains the unusual gate dependence of carrier mobility, spin-orbit coupling, magnetism and superconductivity in these systems. However, progress toward this goal will go beyond electrostatic control of the carrier density and will require careful engineering of the well shape using optimized growth and doping techniques. 

% \section{Acknowledgements}
The authors acknowledge financial support through the DESCO program of the Foundation for Fundamental Research on Matter (FOM), associated with the Dutch Organization for Scientific Research (NWO), and the European Research Council (ERC) through a Consolidator Grant.

\bibliography{Interfaces_in_Complex_Oxides}

\end{document}